\documentclass[useAMS,usenatbib]{mn2e}
\usepackage[utf8]{inputenc}

\usepackage[english]{babel}

\def\shorttitle{Outflows with misaligned magnetic fields}
\def\mytitle{Smoothed particle magnetohydrodynamic simulations of protostellar outflows with misaligned magnetic field and rotation axes}
\def\myauthor{B. T. Lewis, M. R. Bate, and D. J. Price}

\usepackage{graphicx}

\usepackage{aas_macros}
\usepackage{amsmath}
\usepackage{amssymb}
\usepackage{textcomp}
\usepackage{gensymb}
\usepackage{empheq}
\usepackage{csquotes}
\usepackage[hidelinks, pdfauthor={\myauthor}, pdftitle={\mytitle}, pagebackref=false]{hyperref}
\usepackage{cleveref}

\usepackage{setspace}

\usepackage{color,soul} 
\def\diff{} 

\title[\shorttitle{}]{\mytitle{}}
\author[\myauthor{}]{Benjamin T. Lewis$^{1,2}$\thanks{E-mail:
blewis@astro.ex.ac.uk}, Matthew R. Bate$^{1,2}$, and Daniel J. Price$^{2}$\\
$^{1}$School of Physics and Astronomy, University of Exeter, Stocker Road, Exeter EX4 4QL\\
$^{2}$Monash Centre for Astrophysics, School of Mathematical Sciences, Monash University, Clayton, Vic 3800, Australia}

\begin{document}

\date{Accepted 2015 April 28. Received 2015 April 23; in original form 2015 March 06.}

\pagerange{\pageref{firstpage}--\pageref{lastpage}} \pubyear{2015}

\maketitle

\label{firstpage}

\begin{abstract}
We have developed a modified form of the equations of smoothed particle magnetohydrodynamics which are stable in the presence of very steep density gradients. 
Using this formalism, we have performed simulations of the collapse of magnetised molecular cloud cores to form protostars and drive outflows. Our stable formalism
allows for smaller sink particles ($<$ 5 AU) than used previously and the investigation of the effect of varying the angle, $\vartheta{}$, between the initial field axis and the 
rotation axis. 
The nature of the outflows depends strongly on this angle: jet-like outflows are not produced at all when $\vartheta > 30\degree$, and a collimated outflow is not sustained when $\vartheta > 10\degree$. No substantial outflows of any kind are produced when $\vartheta > 60\degree$. This may place constraints on the geometry of the magnetic field in molecular clouds where bipolar outflows are seen. 
\end{abstract}


\section{Introduction}
\label{sec:intro}
Magnetic fields are one of the most important forces influencing the formation of protostars and may resolve several questions about the formation of stars that are left unanswered by purely hydrodynamic theories. That the molecular clouds that ultimately produce protostars are magnetised is well known \citep{1993ApJ...407..175C,2012ARA&A..50...29C}, and this can be confirmed with observations of Herbig-Haro objects with distinctive bipolar outflows, which must have a magnetic origin. Additionally, these outflows may help explain the difference between the angular momentum observed in molecular cloud cores and that of the resultant stars. This is, however, dependent on the ability of the protostar to produce a strong outflow -- if at certain angles this is impossible this may place constraints on the initial field geometry. Recent advances in observational technology have shown that this magnetic field structure can be quite complex \citep{2014Natur.514..597S} and that the hitherto common assumption that field, outflow, and rotation axis are all aligned may be incorrect \citep{2013ApJ...768..159H}. \diff{On very small scales} \citet{2010MNRAS.409.1347D} \diff{have observed a $20\degree$ misalignment between the rotation and field axes of AA Tau.} Previous work, e.g. \citet{2010MNRAS.409L..39C}, using AMR codes has shown that the nature and extent of the protostellar outflow is strongly dependent on the angle (which we denote with $\vartheta$) between the field and rotation axis.  

Smoothed particle hydrodynamics (SPH) methods have been applied to many problems related to the formation of stars, beginning with the original work of \citet{1977AJ.....82.1013L}. These hydrodynamic methods have been extended to magnetohydrodynamics (MHD), originally by \citet{1977MNRAS.181..375G} and \citet{1985MNRAS.216..883P}, with limited success. 
Hitherto, such work has been limited by various numerical instabilities, ranging from unphysical pairing (the `tensile instability') of SPH particles \citep{1995JCoPh.116..123S,2001ApJ...561...82B} to the production of equally unphysical non-solenoidal fields \citep{2012JCoPh.231.7214T}. The most recent instability, and the one that this paper tackles is that a formalism of smoothed particle hydrodynamics incorporating fixes to all of the above deficiencies \citep{2012JCoPh.231..759P} is unstable when `small' sink particles \citep{1995MNRAS.277..362B} are used.

Most recently, \citet{2012MNRAS.423L..45P} examined the collapse of a magnetised molecular cloud core all the way to the formation of the first hydrostatic core \citep{1969MNRAS.145..271L} and \citet{2014MNRAS.437...77B} have continued to the stellar core.
To model the evolution significantly beyond protostar formation, sink particles are a necessary evil since modelling both the magnetohydrodynamics of the protostar and also the surrounding cloud is computationally unfeasible due to the widely different length and time scales involved. Consequently, some way of stabilising the equations of smoothed particle magnetohydrodynamics (SPMHD) in these cases is essential to make progress. These instabilities seem to be only magnified by mis-aligned fields. Previous SPMHD modelling of collapsing cores use somewhat large (5 AU) sink particles and thus, whilst stable, failed to capture the full range of physics. As well as providing a deeper understanding of the formation of individual stars, such a formalism could then be used in larger, cluster size, simulations similar to \citet{2012MNRAS.419.3115B} but with the addition of magnetic fields. Previous cluster-scale simulations performed with SPMHD used the Euler potential method which is limited to certain field geometries \citep{2008MNRAS.385.1820P,2009MNRAS.398...33P}. Given the huge range of density present in following a magnetised cloud collapse, a Lagrangian method such as SPH is ideal. Using our modified SPMHD method we are able to follow the collapse much further than previously, with arbitrarily small sink particles, and at a much higher resolution. 

In \cref{sec:method} we describe our SPMHD formalism, the cause of the instability seen in previous work, and the modifications made to eliminate this. \Cref{sec:init} details our initial conditions. We then perform a low resolution test using a differentially rotating `accretion disc' and also a collapsing magnetised cloud core to demonstrate this modification in \cref{sec:algor}. Finally, in \cref{sec:res} we apply this new formalism to the collapse of magnetised cloud cores with several different values of $\vartheta$ and we discuss the effects of varying this parameter.

\section{Method}
\label{sec:method}

\subsection{Standard SPMHD}

As in \citet{2012MNRAS.423L..45P}, we evolve the equations of ideal magnetohydrodynamics 
with the addition of gravity, viz.
\begin{equation}
\frac{\rmn{d}~}{\rmn{d}t}\rho{} = -\rho{}\nabla{}^{i}v^{i} ~\text{,}
\end{equation}
\begin{equation}
\frac{\rmn{d}~}{\rmn{d}t} v^{i} = \frac{1}{\rho} S^{ij} - \nabla^{i}\phi{}
\label{eqn:mom_analyitc}
\end{equation}
\begin{equation}
\frac{\rmn{d}~}{\rmn{d}t} B^{i} = \left(B^{j}\nabla^{j} \right)v^{i} - B^{i} \left(\nabla^{j} v^{j} \right) ~\text{,}
\end{equation}
\begin{equation}
\nabla^{2}\phi{} = 4\pi{}G\rho{} ~\text{,}
\end{equation}
with the MHD stress tensor is given by
\begin{equation}
S^{ij} = - P \delta^{ij} +\frac{1}{\mu_0} \left( B^i B^j - \frac{1}{2}\delta^{ij}B^2 \right)
\end{equation}
and where, as usual, $\rho{}$, $v^{i}$, $B^{i}$, $P$, $\phi{}$ represent density, velocity, the magnetic field strength, the 
hydrodynamic pressure and the gravitational potential, respectively; repeated indicies imply summation, and
\begin{equation} 
\frac{\rmn{d}~}{\rmn{d}t} = \frac{\partial{}~}{\partial{}t} + v^{i}\nabla^{i} ~\text{.}
\end{equation}

We evolve these equations using the method of smoothed particle magnetohydrodynamics described in \citet{2005MNRAS.364..384P} and \citet{2012JCoPh.231..759P}
and artificial viscosity and resistivity terms based on Riemann solvers \citep{1997JCoPh.136..298M} with temporal and spatially dependent switches. We use the \citet{1997JCoPh.136...41M} switch for artificial viscosity, with $\alpha_{\rmn{AV}} \in [0.1, 1.0]$ and the newer \citet{2013MNRAS.436.2810T} switch for artificial resistivity with $\alpha_{\rmn{B}} \in [0.0, 1.0]$. This differs slightly from \citet{2012MNRAS.423L..45P} which used an older resistive switch and constrained $\alpha_{\rmn{B}}$ to  $[0.0, 0.1]$. We soften the gravitational potential using the same SPH smoothing kernel as used in the rest of the simulation \citep{2007MNRAS.374.1347P}.

All magnetic fields are solenoidal and consequently we must maintain a divergence free field. This is not naturally satisfied in SPMHD. Consequently we use the constrained hyperbolic divergence cleaning of \citet{2012JCoPh.231.7214T} (which is based on the earlier \citet{2002JCoPh.175..645D} method used in some grid codes). With this, we have an additional scalar field, $\psi$ in the induction equation such that:
\begin{equation}
\frac{\rmn{d}~}{\rmn{d}t} B^{i}|_{\rmn{clean}} = - \nabla^{i}\psi ~\text{,}
\end{equation}
where
\begin{equation}
\frac{\rmn{d}~}{\rmn{d}t}\psi{} = - c^{2}_{c} \nabla^{i}B^{i} - \frac{\psi}{\tau} - \frac{1}{2}\psi{}\left( \nabla^{i}v^{i} \right) ~\text{.}
\end{equation}
This removes the unphysical divergence by propagating a damped wave through the simulation. Unlike in \citet{2014MNRAS.437...77B}, we do not need to increase the cleaning wave, $c_c$, speed above the magnetosonic speed to maintain stability, avoiding a costly decrease in the size of the timesteps. We set the damping timescale, 
\begin{equation}
\tau = \frac{h}{\sigma{}c_c} ~\text{,}
\end{equation}
to be critically damped with $\sigma = 0.8$, where $h$ is the SPH smoothing length. 

As in \citet{2004MNRAS.348..139P}, we use a variable smoothing length formalism to ensure that computational resources are used efficently and that sufficent resolution is applied to the complicated areas of the model. The smoothing length and density are solved self-consistently (via the Newton-Raphson iterative method) using
\begin{equation}
h = \eta{}\left( \frac{m}{\rho} \right)^{\frac{1}{\nu{}}} ~\text{,}
\label{eqn:h-rho}
\end{equation}
where $\eta_{} = 1.2$ for the `standard' cubic B-spline kernel \citep{1985JCoPh..60..253M}, and $\nu = 3$ is the number of spatial dimensions.

The simulations were performed using a three-dimensional SPH code -- originally written by \citet{1990ApJ...348..647B} but extensively modified by Bate and his collaborators \citep{2009MNRAS.392..590B} -- with the additional modifications detailed in the next section. The code uses a binary tree to both find neighbours for particles and to calculate gravity. A second--order Runge-Kutta-Fehlberg integrator \citep{RK4NASATR} with each particle carrying an individual timestep \citep{1995MNRAS.277..362B} was used to evolve the simulation in time.
\diff{Each simulation was run on a single 12-core hyperthreaded compute node (i.e. 24 execution threads in total), taking between 400 hours of wall time (4,500 core hours) for the simpler aligned models to over 550 wall hours (6,600 core hours) for  the more complicated highly misaligned models.}

\subsection{The `average h' method}

As noted earlier, an SPMHD instability exists in regions where the density gradient is very large. 
Since $\rho{}$ and the smoothing length, $h$ are related by \cref{eqn:h-rho} any density gradient will naturally produce
an inverse gradient in the smoothing length. For conventional SPH, and indeed for SPMHD where the gradients are more gentle, 
this does not produce any stability problems. To discover why this is an issue, we consider the correction to the tensile
instability alluded to earlier.

Separating the MHD stress tensor into two parts (for the remainder of this paper we
will ignore the gravitational potential terms) such that
\begin{equation}
S^{ij} = S^{ij}|_{\rm{isotropic}} + S^{ij}|_{\rmn{anisotropic}} \text{~,}
\end{equation}
where
\begin{equation}
S^{ij}|_{\rm{iso}} = \left( -P + \frac{1}{2}\frac{1}{\mu_{0}} B^{2} \right) \delta^{ij} \text{~,}
\end{equation}
\begin{equation}
S^{ij}|_{\rm{anis}} = \frac{1}{\mu_{0}} B^{i}B^{j} \text{~,}
\end{equation}
we observe that the anisotropic component of the momentum equation (\cref{eqn:mom_analyitc}) can be written as
\begin{align}
\begin{split}
\frac{\rmn{d}~}{\rmn{d}t} v^{i}|_{\rm{anis}} & = - \frac{1}{\rho{}}\nabla^{j}S^{ij}|_{\rmn{anis}} = - \frac{1}{\rho{}}\frac{1}{\mu_{0}}\nabla^{j} B^{i}B^{j} \\
& = -\frac{1}{\rho{}}\frac{1}{\mu_{0}} \left[ \left(B^{j}\nabla^{j}\right)B^{i} + B^{i}\left(\nabla^{j}B^{j}\right)\right] \text{~,}
\end{split}
\end{align}
where we note the important constraint -- $\nabla{}\cdot{}$\mbox{\boldmath{$B$}}= $0$ -- that implies that the $B^{i}\left(\nabla^{j}B^{j}\right)$ term must be zero to ensure a solenoidal field. As noted before this is not true in general in SPH and an unphysical force along the field lines may be produced -- this is the `tensile instability' \citep{1995JCoPh.116..123S}. Specifically, this results in the stress tensor becoming positive, and hence an attractive force between particles being generated. 
The cleaning described earlier is correcting for a different manifestation of the same problem, and hence will not help here.  The correction proposed by \citet{2001ApJ...561...82B} is to subtract 
a source term which is exactly equal to this unphysical divergence. In our formulation of SPMHD we use a symmetric operator for the anisotropic component of the momentum equation, viz.
\begin{align}
\begin{split}
\frac{\rmn{d}~}{\rmn{d}t} v^{i}_a|_{\rmn{anis,full}} = \frac{1}{\mu_{0}} &\sum^{N}_{b} m_{b} \left( \frac{B^{i}_{a}B^{j}_{a}}{\Omega_{a}\rho^{2}_{a}}\nabla^{j}_{a}W_{ab}(h_{a}) \right. \\
&+ \left. \vphantom{\sum^{N}_{b}}\frac{B^{i}_{b}B^{j}_{b}}{\Omega_{b}\rho^{2}_{b}}\nabla^{j}_{a}W_{ab}(h_{b}) \right) \text{~,}
\end{split}
\end{align}
where $W_{ab}\left( h_{\left\{a,b\right\}} \right) = W\left(r^{i}_{a} - r^{i}_{b}, h_{\left\{a,b\right\}} \right)$ is the smoothing kernel, $a$ and $b$ represent individual SPH particles, $N$ is the number of neighbour particles (i.e. particles for which $W_{ab} \ne 0$),
\begin{align}
\begin{split}
\Omega_{a} &= 1 - \frac{\partial{}h_{a}}{\partial{}\rho_{a}} \sum^{N}_{b} \frac{\partial{}W_{ab}\left( h_a\right) }{\partial{}h_{a}}\\
&= 1 + \frac{h_{a}}{\nu{}\rho_{a}}\sum^{N}_{b}\frac{\partial{}W_{ab} \left( h_{a} \right) }{\partial{}h_{a}}\text{~,}
\end{split}
\end{align}
are terms to take account of gradients in $h$, $\nu$ is the number of spatial dimensions, and all other terms have the usual meanings. Consequently, we use a symmetric
operator to estimate the magnetic divergence, 
\begin{align}
\begin{split}
\nabla^{j}_{a}B^{j}_{a} = \frac{1}{\mu{}_{0}} &\sum^{N}_{b} m_{b} \left( \frac{B^{j}_{a}}{\Omega_{a}\rho^{2}_{a}} \nabla^{j}_{a}W_{ab}\left(h_{a}\right) \right. \\
&+ \left. \vphantom{\sum^{N}_{b}} \frac{B^{j}_{b}}{\Omega_{b}\rho^{2}_{b}} \nabla^{j}_{a}W_{ab}\left(h_{b}\right) \right) \text{~,}
\label{eqn:divb}
\end{split}
\end{align}
which is then subtracted from the momentum equation such that,
\begin{equation}
\frac{\rmn{d}~}{\rmn{d}t} v^{i}|_{\rmn{anis}} = \frac{\rmn{d}}{\rmn{d}t} v^{i}|_{\rmn{anis,full}} - \chi{}B^{i}_{a}\left(\nabla^{j}_{a}B^{j}_{a}\right) \text{~.}
\label{eqn:divbsub}
\end{equation}
We use $\chi{} = 1$ as recommended by \citet{2012JCoPh.231.7214T}. Whilst this necessarily makes the equation non-conservative, it is much more stable than the value
of $\frac{1}{2}$ recommended by \citet{2001ApJ...561...82B}. 
This means the anisotropic momentum equation we evolve is given by 
\begin{equation}
\frac{\rmn{d}~}{\rmn{d}t} v^{i}_a|_{\rmn{anis}} = \frac{1}{\mu{}_{0}} \sum^{N}_{b} \frac{m_{b}}{\Omega_{b}\rho^{2}_{b}}\left(B^{i}_{b} - B^{i}_{a} \right)B^{j}_{b}\nabla{}^{j}_{a}W_{ab}\left(h_{b}\right) \text{~.}
\label{eqn:spmhdaniso}
\end{equation}
Comparing this to the SPMHD induction equation, 
\begin{equation}
\hspace*{-9pt}
\frac{\rmn{d}~}{\rmn{d}t} \left( \frac{B^{i}_{a}}{\rho_{a}} \right) = -\frac{1}{\Omega_{a}\rho^{2}_{a}} \sum^{N}_{b} m_{b} \left( v^{i}_{a} - v^{i}_{b} \right) B^{j}_{a} \nabla^{j}_{a}W_{ab}\left(h_{a}
\right) \text{,}
\label{eqn:spmhdind}
\end{equation}
we observe that \cref{eqn:spmhdind} depends only on $h_a$ and \cref{eqn:spmhdaniso} only upon $h_b$. In situations where the gradients in $\rho$ are small, this does not present any major issues. However, if $\rho_{a} \gg \rho_{b}$ 
then $h_{a} \ll h_{b}$, consequently, it is possible for some particle $a$ to have the temporal evolution of its magnetic field evaluated over a very small number of neighbours but a force from that field interpolated over a
large number. This is clearly undesirable, and is the cause of the violent instabilities seen in many previous SPMHD calculations with large density gradients. To resolve this, we replace the $h_{\left\{a,b\right\}}$ terms in these two equations with
\begin{equation}
\bar{h}_{ab} = \frac{1}{2}\left( h_{a} + h_{b} \right) \text{~,}
\end{equation}
such that
\begin{equation}
\frac{\rmn{d}~}{\rmn{d}t} v^{i}_{a}|_{\rmn{anis}} = \frac{1}{\mu{}_{0}} \sum^{N}_{b} \frac{m_{b}}{\rho^{2}_{b}}\left(B^{i}_{b} - B^{i}_{a} \right)B^{j}_{b}\nabla{}^{j}_{a}W_{ab}\left(\bar{h}_{ab}\right) \text{~,}
\label{eqn:avhmom}
\end{equation}
and for the SPMHD induction equation, 
\begin{equation}
\frac{\rmn{d}~}{\rmn{d}t} \left( \frac{B^{i}_{a}}{\rho_{a}} \right) = -\frac{1}{\rho^{2}_{a}} \sum^{N}_{b} m_{b} \left( v^{i}_{a} - v^{i}_{b} \right) B^{j}_{a} \nabla^{j}_{a}W_{ab}\left(\bar{h}_{ab}\right) \text{~.}
\label{eqn:avhind}
\end{equation}
This corrects the instability since in the limit $h_{b} \ll h_{a}$ $\bar{h}_{ab} \rightarrow h_{a}$ and vice versa. The stability seen in \citet{2012MNRAS.423L..45P} was simply a product of the larger sink radii preventing the formation of density gradients so extreme that this is an issue. Similarly, in the outer regions of the collapse simulation where the density gradient is much flatter it does not cause a large change in the smoothing length, preventing an undesirable loss of resolution in these areas. 
Whilst the subtraction in \cref{eqn:divbsub} makes the equations of SPMHD less conservative; \cref{eqn:divb} derives entirely from \cref{eqn:spmhdind} and consequently no additional conservation loss is introduced by the use of the average h terms.
However, the removal of the $\Omega$ terms from \cref{eqn:spmhdaniso} and \cref{eqn:spmhdind} may introduce a very small error.

Since all other SPH equations (except the density/smoothing length) contain terms for both $h_a$ and $h_b$ it is not necessary to apply this correction elsewhere. In particular, we do not apply it to the hyperbolic cleaning terms. Even though the density is evaluated using only one smoothing length, this does not contribute to the instability since, in effect, $\rho$ and $h$ are the same parameter and are therefore consistent. We did investigate only applying the average \textit{h} method to only one of \cref{eqn:avhmom} or \cref{eqn:avhind}, and observed that this was less stable that applying it to both -- which is the expected result given that $\bar{h}_{ab}$ will tend towards the larger value.

We also considered a slightly simpler scheme, whereby we imposed a minimum smoothing length, $h_{\rmn{min}}$ on the whole simulation by modifying \cref{eqn:h-rho} to be
\begin{equation}
h_{a}(r_{a}) = \eta{}\left( \frac{m_{a}}{\rho_{a}} \right)^{\frac{1}{\nu{}}} + h_{\rmn{min}} \text{,}
\end{equation}
which has the desirable properties that it does not introduce any extra loss of conservation and does not change the $\Omega$ terms (since there is no spatial dependence to $h_{\rmn{min}}$). However, it is difficult to determine a `correct value' of $h_{\rmn{min}}$ a priori. In addition this formalism would cause benign particle pairing (see \citet{2012JCoPh.231..759P} for details) due to particles having too many neighbours as well as needlessly sacrificing resolution on other SPH equations. We found that the only effective values of $h_{\rmn{min}}$ were so large that the loss of resolution caused by pairing was in itself a serious problem.

We also investigated, unsuccessfully, using the average of two smoothing kernels, i.e.
\begin{equation}
\overline{W}_{ab} = \frac{1}{2}\left( W_{ab}\left(h_{a}\right) + W_{ab}\left(h_{b}\right) \right) \text{~,}
\label{eqn:avkern}
\end{equation}
which was little different to the status quo.   
In the limit where $h_a >> h_b$ then $W_{ab}(h_a) >> W_{ab}(h_b)$. This will result in the average in \cref{eqn:avkern} essentially becoming $\frac{1}{2}W_{ab}(h_{a})$. Whilst this would be desirable for one of the two MHD equations, it will reduce the smoothing applied to the other substantially. Consequently this approach was rejected. 

\section{Initial Conditions}
\label{sec:init}

The initial conditions for our calculations of protostellar collapse are broadly the same as those in \citet{2012MNRAS.423L..45P}. However, we use more SPH particles and smaller accretion radii for our sink particles.
We begin with a 1.5 million SPH particle uniform density sphere of cold gas, more than sufficient to resolve a Jeans length according to the criteria in \citet{1997MNRAS.288.1060B}, placed in a periodic box and surrounded by an external medium of ca. 500,000 warm gas particles.
There is a density ratio of 30:1 between the warm outer medium and the cool sphere with a pressure equilibrium between the sphere and the medium. Particles are initially laid out on a cubic lattice, the initial radius of the sphere is $r_{\rmn{cloud}} = 4\times{}10^{15}~\rmn{cm}$ with a mass of $M = 1 \rmn{M}_{\odot}$
giving an initial density in the sphere $\rho_{0} = 7.4\times{}10^{-18}~\rmn{g}~\rmn{cm^{-3}}$
The sphere has an initial isothermal sound speed $c_{s} = 2.2\times{}10^{4}~\rmn{cm}~\rmn{s}^{-1}$ and we use the barotropic equation of state

\begin{empheq}[left={P = c^{2}_{s} \empheqlbrace}]{equation}
\begin{aligned}
& \rho{}          &&\rho{} \leq \rho_{\rmn{c,1}} \\
& \rho_{\rmn{c,1}} \left( \frac{\rho{}}{\rho_{\rmn{c,1}}} \right)^{\frac{7}{5}} & \hspace{-5pt}\rho_{\rmn{c,1}} <~ & \rho{} \leq \rho_{\rmn{c,2}} \\
& \rho_{\rmn{c,1}} \left( \frac{\rho_{\rmn{c,2}}}{\rho_{\rmn{c,1}}} \right)^{\frac{7}{5}} \rho_{\rmn{c,2}}\left( \frac{\rho{}}{\rho_{\rmn{c,2}}} \right)^{\frac{11}{10}} && \rho{} > \rho_{\rmn{c,2}}
\end{aligned}
\end{empheq}
where the two critical densities are given by $\rho_{\rmn{c,1}} = 10^{-14}~\rmn{g}~\rmn{cm}^{-3}$ and $\rho_{\rmn{c,2}} = 10^{-10}~\rmn{g}~\rmn{cm}^{-3}$. This is similar to that used, for example, in \citet{2008ApJ...676.1088M} with the removal of the final $\gamma = \frac{5}{3}$ step at the highest densities. The sphere has an initial temperature of approximately $10~\rmn{K}$; since the outer medium also begins with the same initial pressure it has a correspondingly higher initial temperature of approximately $300~\rmn{K}$. 
The sphere is set in solid body rotation at $\Omega = 1.77\times{}10^{-13}~\rmn{rad}~\rmn{s}^{-1}$, such that the magnitude of the ratio of rotational to gravitational energy is $\approx{} 0.005$, within the range observed by \citet{1993ApJ...406..528G}.

We then define a new parameter, $\vartheta{}$, which is the angle between the rotation axis of the sphere (which is always aligned with the z-axis)
and the initial magnetic field. The magnetic field is then initially
\begin{align}
B_{x} = B_{0}~\rmn{sin}~\vartheta{} \text{~,}\\
B_{z} = B_{0}~\rmn{cos}~\vartheta{} \text{~,}
\end{align}
i.e. when $\vartheta{} = 0 \degree{}$ the field is aligned with the z-axis. 
The initial magnetic field $B_0$ is determined using the parameter $\mu$, which is \citep{1978PASJ...30..671N,2004RvMP...76..125M} the ratio between the sphere's mass-flux ratio and the critical mass-flux ratio for a spherical cloud, i.e.
\begin{equation}
\mu{} = \frac{\mu_{\rmn{cloud}}}{\mu_{\rmn{crit}}} \text{~,}
\end{equation}
where,
\begin{equation}
\mu_{\rmn{cloud}} = \frac{M}{\pi{}r^{2}_{\rmn{cloud}}B_{0}} \text{,}\quad{} \mu_{\rmn{crit}} = \frac{2c_{1}}{3}\sqrt{\frac{5}{\pi{}G\mu_{0}}} \text{~,}
\end{equation}
with the ratio between the minimum self-collapsing gravitational mass obtained from the virial theorem and that required for a magnetised astrophysical cloud, $c_1 = 0.53$ as obtained numerically by \citet{1976ApJ...210..326M}. Throughout this paper we use $\mu = 5$.

Sink particles are added to the simulation once a critical density of $10^{-10}~\rmn{g}~\rmn{cm^{-3}}$ is achieved, and the usual tests are passed \citep{1995MNRAS.277..362B}. Our more stable formalism allows for smaller sink sizes that used in \citet{2012MNRAS.423L..45P}; we use an accretion radius of 1 AU as a compromise between capturing physics and numerical efficiency. Tests have been performed with sink particles with accretion radii of 0.1 AU and 0.01 AU. Our sink particle will accrete unconditionally once a particle crosses its accretion radius; since all our simulations are of a collapsing core this should not result in any deleterious effects. As in previous work, the sink particle does not carry a magnetic field -- when a particle is eliminated from the simulation, the mass is added to the sink (which does not exert a hydrodynamic pressure). 

\section{Algorithm Tests}
\label{sec:algor}

\subsection{Isothermal cylinder in a box}

\begin{figure}
\centering{}
\includegraphics{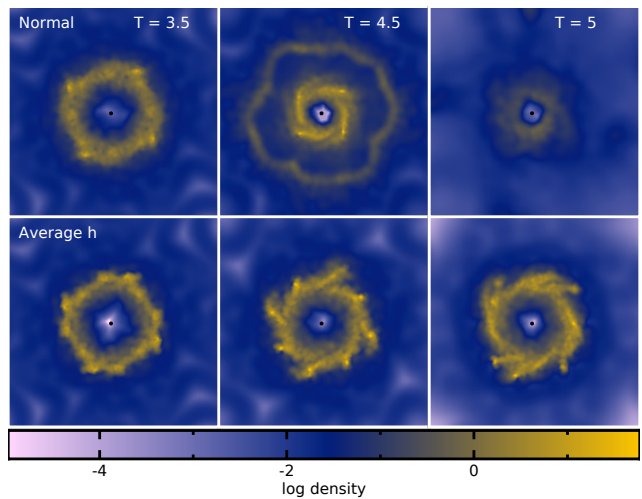}

\caption{Density cross-sections in the $x-y$ plane for a differentially rotating disc. The top row is the `standard' SPMHD formalism whilst the bottom is the average $h$ method. The unphysical bubble caused by the instability discussed above can be clearly seen at $t$ = 4.5.\label{fig:accdiscrho}}
\end{figure}

\begin{figure}
\centering{}
\includegraphics{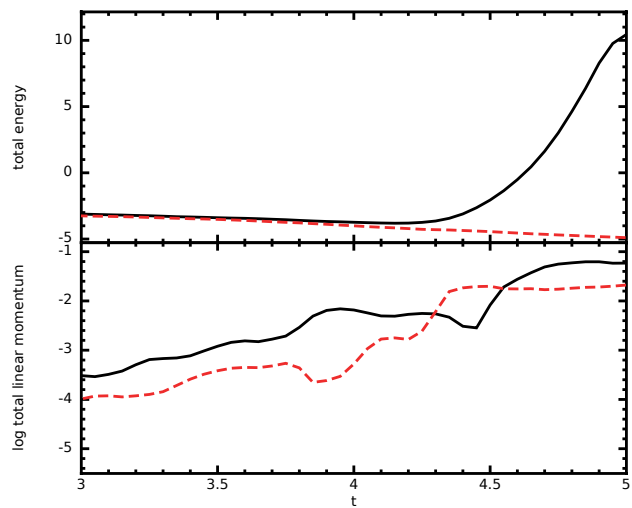}

\caption{Total energy and total linear momentum for the `standard' SPMHD method (solid line) and the average-h method (dashed line). The rapid increase in the total energy is correlated with the unphysical bubble seen in \cref{fig:accdiscrho}. 
\label{fig:accdiscgrab}}
\end{figure}

In addition to performing full-scale models, we performed a series of low resolution tests. An 8,000 SPH particle isothermal cylinder was created inside a periodic box, with a central potential provided by a sink particle with a mass 10 times that of the cylinder. We use a sink particle rather than a potential well to prevent a high density region centred on the origin requiring a very short time-step. The inner and outer cylinder radii were 0.5 and 5 code units respectively and the cylinder was 2.5 units thick (i.e. a height to radius ratio of $\frac{1}{2}$), and the sink particle had an accretion radius of 0.3 code units. The cylinder was then set in differential rotation, with a $r^{-2}$ velocity profile. The initial velocity was set such that the rotation period at 1 unit distance was $T = 2$. 

A uniform initial sound speed of 0.1 code units was set with an isothermal equation of state $P(\rho) = \frac{2}{3}u\rho{}$; with an initial ratio between hydrodynamic and magnetic pressure (the `plasma $\beta$') of $\beta \approx{} 8.4$ (corresponding to a nominal mass-flux ratio of 5, though this is not a useful measure for a differentially rotating cylinder where the effect of magnetic braking is much more dominant than magnetic pressure). The system was then allowed to evolve. In a correct model, we would expect the cylinder to pile up, with material from within 1 unit radius moving outwards and more distant material spiraling inwards. Some material will fall towards the central sink (both due to magnetic and viscous braking) and be accreted. The cylinder will also flatten due to rotational forces and self gravity, further increasing the density.

\Cref{fig:accdiscrho} shows cross-sections in the $x-y$ plane for the normal SPMHD formalism and our modified one. A clearly unphysical bubble like structure can be seen for the normal code which is not present in our modified method. In the original method, the total energy increases rapidly and quickly becomes positive; in contrast the average $h$ method results in a monotonically decreasing energy, as expected for an isothermal equation of state. Earlier, we noted that this formalism could, in principle, exhibit poorer momentum conservation than the standard form, however, this is in practice undetectable, as seen in \cref{fig:accdiscgrab}.

\subsection{Low resolution spherical collapse}

\begin{figure}
\centering{}

\includegraphics{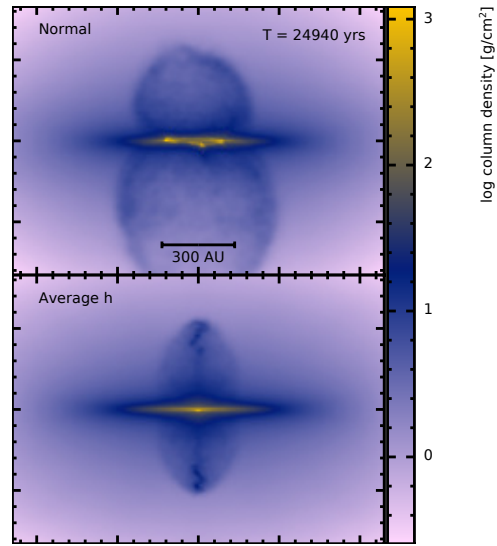}
\caption{Column density plots for the low resolution test with the unmodified scheme (top) and our modified average-h scheme (bottom) at $t$ = 24940 yrs. The large unphysical explosion can be clearly seen in the upper plot, where a large asymmetrical bubble of material has been ejected at very high velocities. In comparison, the modified scheme forms a collimated jet correctly (albeit underresolved due to the low resolution). \label{fig:lowresexp}}
\end{figure}

Using the initial conditions discussed more fully in \cref{sec:init}, we then performed a low resolution (150,000 SPH particles) comparison of the collapse of a spinning magnetised cloud core with $\vartheta{} = 0\degree{}$ using both schemes. \Cref{fig:lowresexp} shows the situation shortly after the insertion of a sink particle. The violent and unphysical explosion in the unmodified code can be clearly contrasted with the symmetric bipolar outflow in the modified code. 

In both cases, the tests for insertion of a sink particle are passed at approximately one free-fall time ($t_{\rmn{ff}} = 24430~\rmn{yrs}$) and the explosion happens at between $1.01~t_{\rmn{ff}}$ and $1.02~t_{\rmn{ff}}$, i.e. soon after the critical density for sink creation is reached.

\begin{figure}
\centering{}

\includegraphics{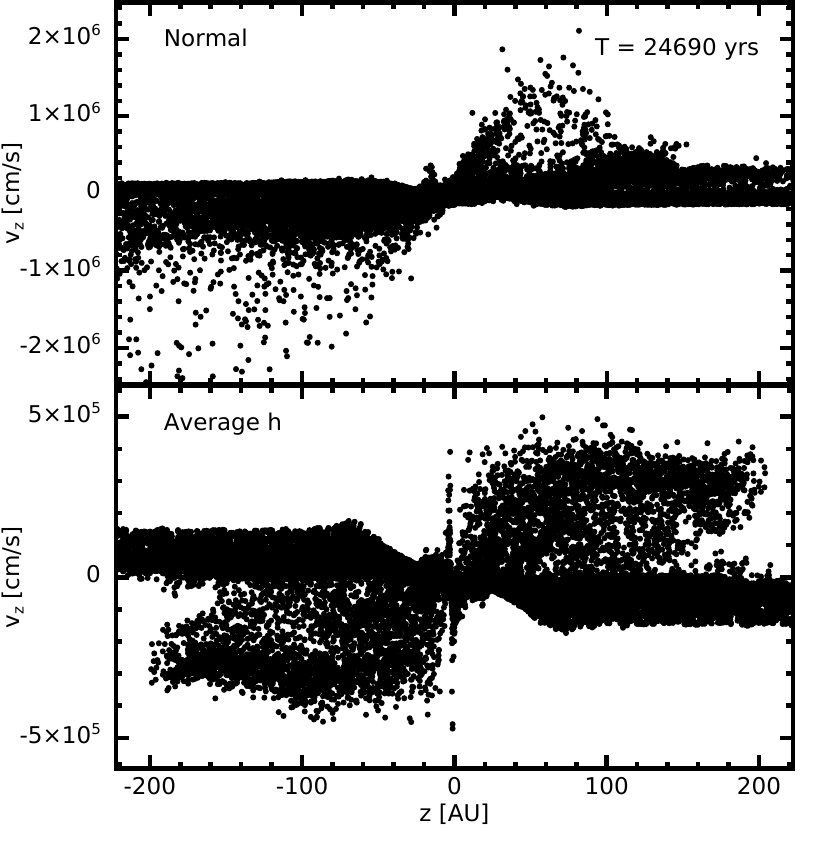}
\caption{Vertical velocity of SPH particles as a function of height for both schemes at the same time as in \cref{fig:lowresexp}. Unlike the modified scheme, the original method does not produce a symmetrical outflow at $\approx{} 5~\rmn{km}~\rmn{s^{-1}}$ (note that this is slower than that seen at higher resolutions later since the accretion region is under-resolved), and the maximum velocities seen are over $40~\rmn{km}~\rmn{s^{-1}}$. Note that the scale on the upper panel is different to that on the lower panel.\label{fig:lowresexpvel}} 
\end{figure}

In the unmodified scheme, a high velocity bubble of material is produced and ejected, similar to that seen in \cref{fig:accdiscrho}, but in this case the most significant effect is to eject material perpendicular to the plane of the disc. This is probably due to this being both the rotation and magnetic field axis, and therefore the preferred direction for momentum transport (similar to how the collimated jet is produced in the modified scheme). 
In \cref{fig:lowresexpvel} the velocity in the vertical as a function of height is shown, which demonstrates the symmetrical and collimated nature of the outflow in the modified code, and the much higher and broadly distributed velocities in the original code. The collimated jet produced by the average $h$ method remains stable until all material has either been accreted or ejected and the jet is extinguished.

\section{Results}
\label{sec:res}

\subsection{Nature of outflows and jets}

\begin{figure*}
\centering{}
\includegraphics{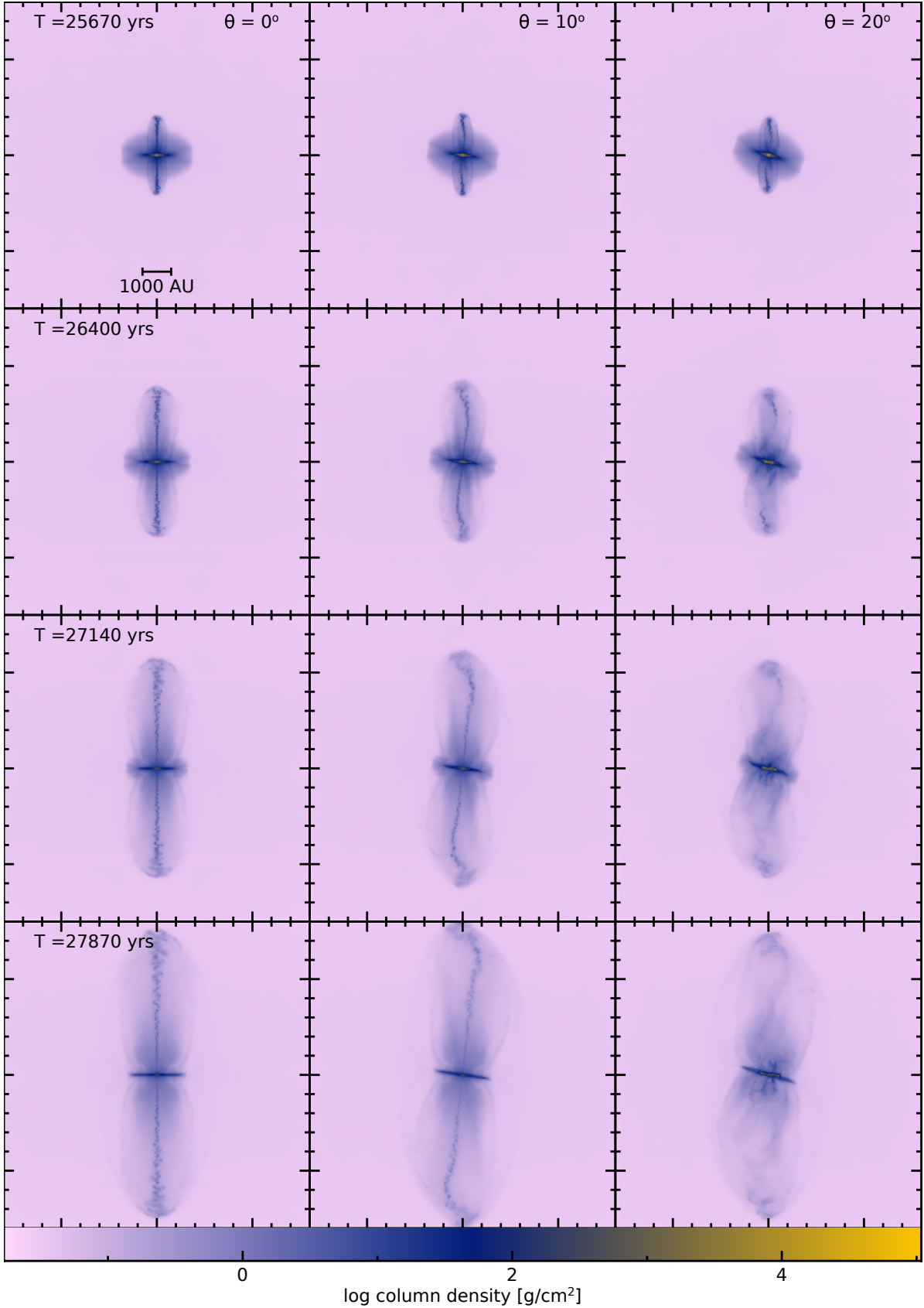}
\caption{Column density projection plots for $\vartheta = [0\degree{},10\degree{},20\degree{}]$ (across the page) at 4 different times (down the page). At these shallow angles, a prominent collimated outflow aligned with the field axis is always formed, however in the $\vartheta = 20\degree$ case this is eventually disrupted and becomes puffy and uncollimated by $t$ = 27870 years. The rotation axis is along the $z$-axis.\label{fig:fig1}} 
\end{figure*}

\begin{figure*}
\centering{}
\includegraphics{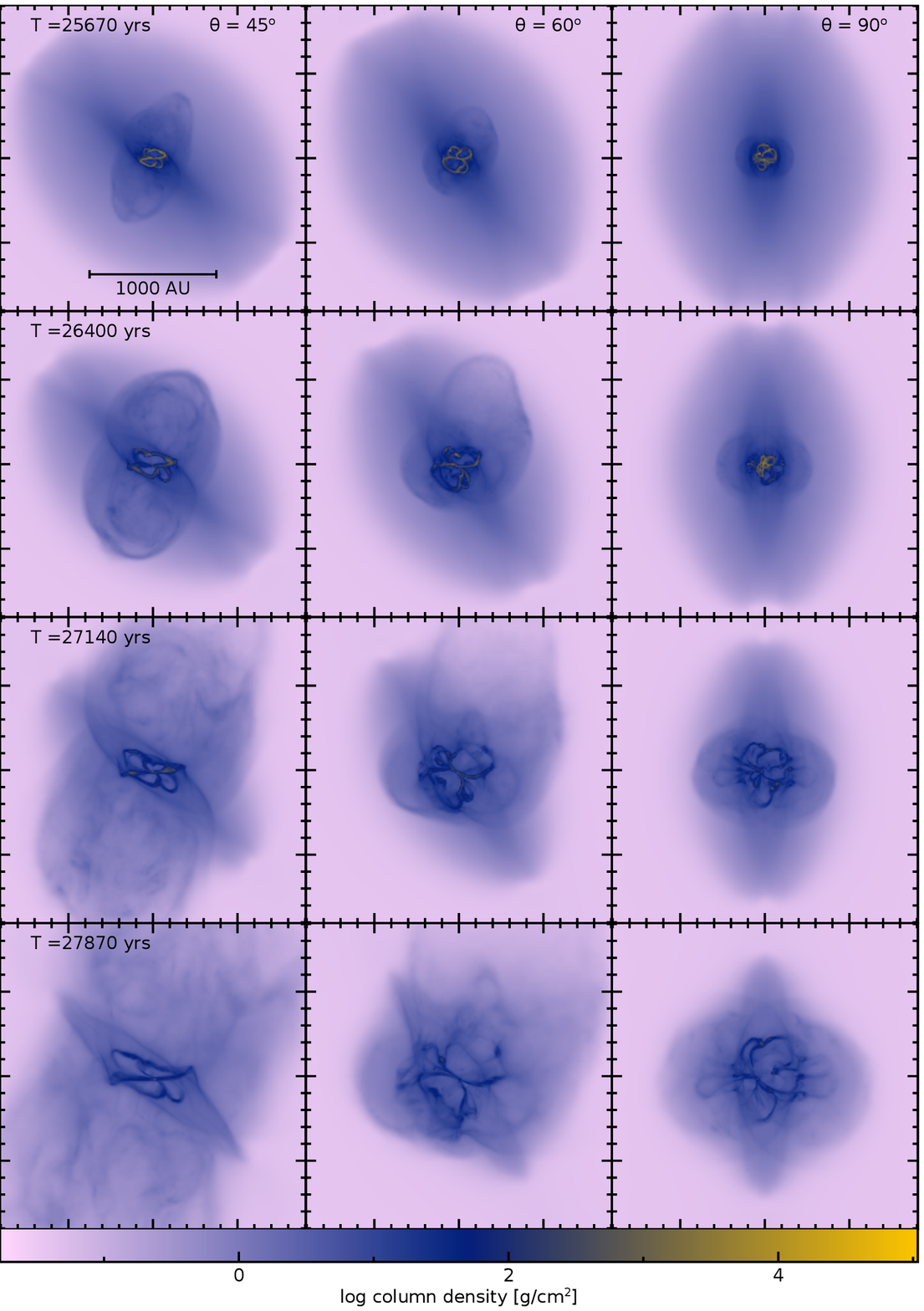}
\caption{Column density projection plots for $\vartheta = [45\degree{},60\degree{},90\degree{}]$ (across the page) at 4 different times (down the page). Note that the scale here is different to \cref{fig:fig1} to show the more complicated inner structures. At these substantially steeper angles no collimated outflow is produced at all and for $\vartheta > 45\degree{}$ the outflow is very heavily suppressed.
\label{fig:fig1a}}
\end{figure*}

We performed calculations with values of $\vartheta = [0\degree{},10\degree{},20\degree{},45\degree{},60\degree{},90\degree{}]$. \Cref{fig:fig1,fig:fig1a} show the time evolution for these six angles. We note that the 
the results for the $\vartheta = 0\degree$ case are broadly the same as in \citet{2012MNRAS.423L..45P}, albeit with a slightly faster jet velocity - in this case $\sim 8~\rmn{km~s^{-1}}$. This is expected, since the smaller accretion radii used here will allow a faster velocity near the sink particle, and since the axial velocity of a collimated jet is proportional to the velocity of the matter spiraling in to create it this naturally leads to a faster jet \citep{2003MNRAS.339.1223P}. This is the only significant difference between this result and the earlier calculation that used a 5 AU sink, showing that our modification to the SPH equations does not cause numerical artifacts or errors of its own.

The most striking result is the lack of any real outflow at all from the $\vartheta = [60,90]\degree$ models. 
Whilst all shallower angles produce an outflow of some significance, this only takes the form of a collimated jet for $\vartheta \leq{} 20\degree{}$, and can only be sustained when $\vartheta{} \leq{} 10\degree{}$. \diff{Similarly, the pseudo-disc which is clearly defined for }$\vartheta = 0\degree$\diff{ is either disrupted or, in the most misaligned cases, does not form at all. Since the formation of a stable bipolar outflow requires a stable and defined disc structure, this naturally prevents a substantial outflow being formed. In the intermediate }$\vartheta = 45\degree$\diff{ case, the pseudo-disc formed is highly disrupted but still manages to drive a broad, albeit slower, outflow.}

\begin{figure}
\centering{}
\includegraphics{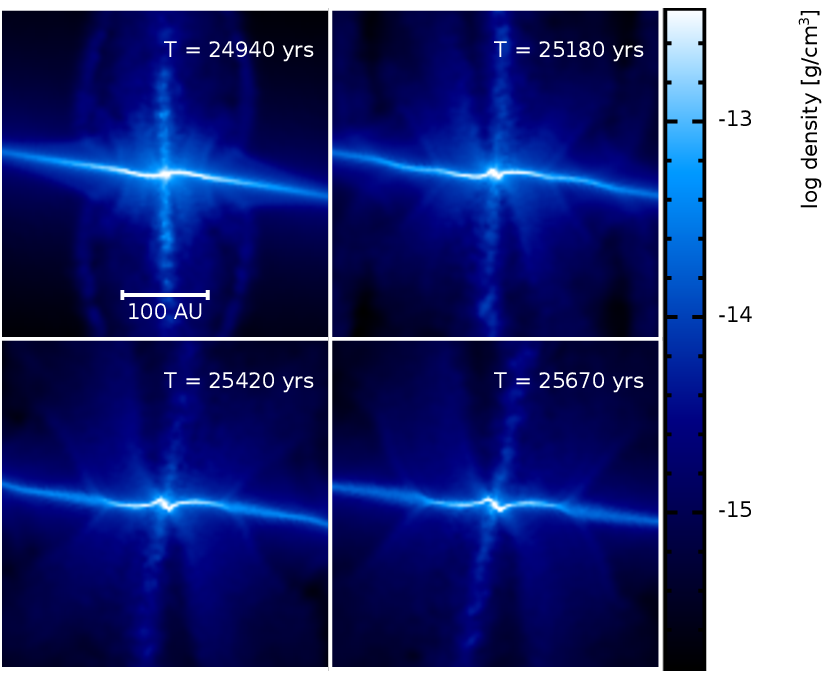}

\caption{Cross-sections of density in the $z-x$ plane for $\vartheta{} = 10\degree{}$. Whilst the outer regions of the pseudo-disc align perpendicular to the magnetic field, the innermost region exhibits a more complicated structure. As it deforms, the collimated jet changes from being parallel to the rotation axis to being parallel to the field axis.\label{fig:10degzoom}}
\end{figure}

\begin{figure}
\centering{}
\includegraphics{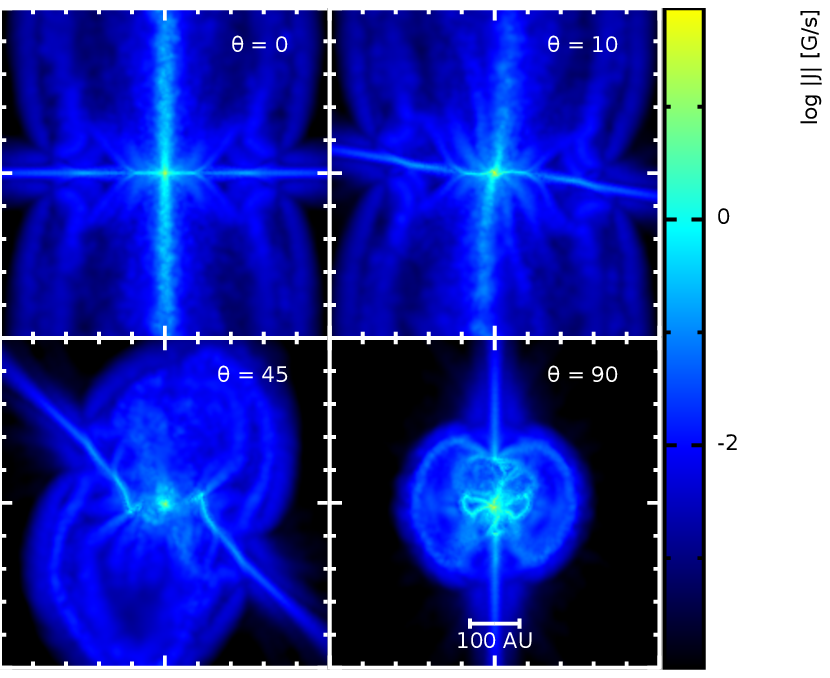}

\caption{Cross-sections of $|\mathbf{J}|$ at $t$ = 25420 yrs for 4 different values of $\vartheta$. The magnetic field geometry is significantly more complicated in the latter 
two cases, and this corresponds with a substantially changed outflow.\label{fig:modJ}}
\end{figure}

\begin{figure}
\centering{}
\includegraphics{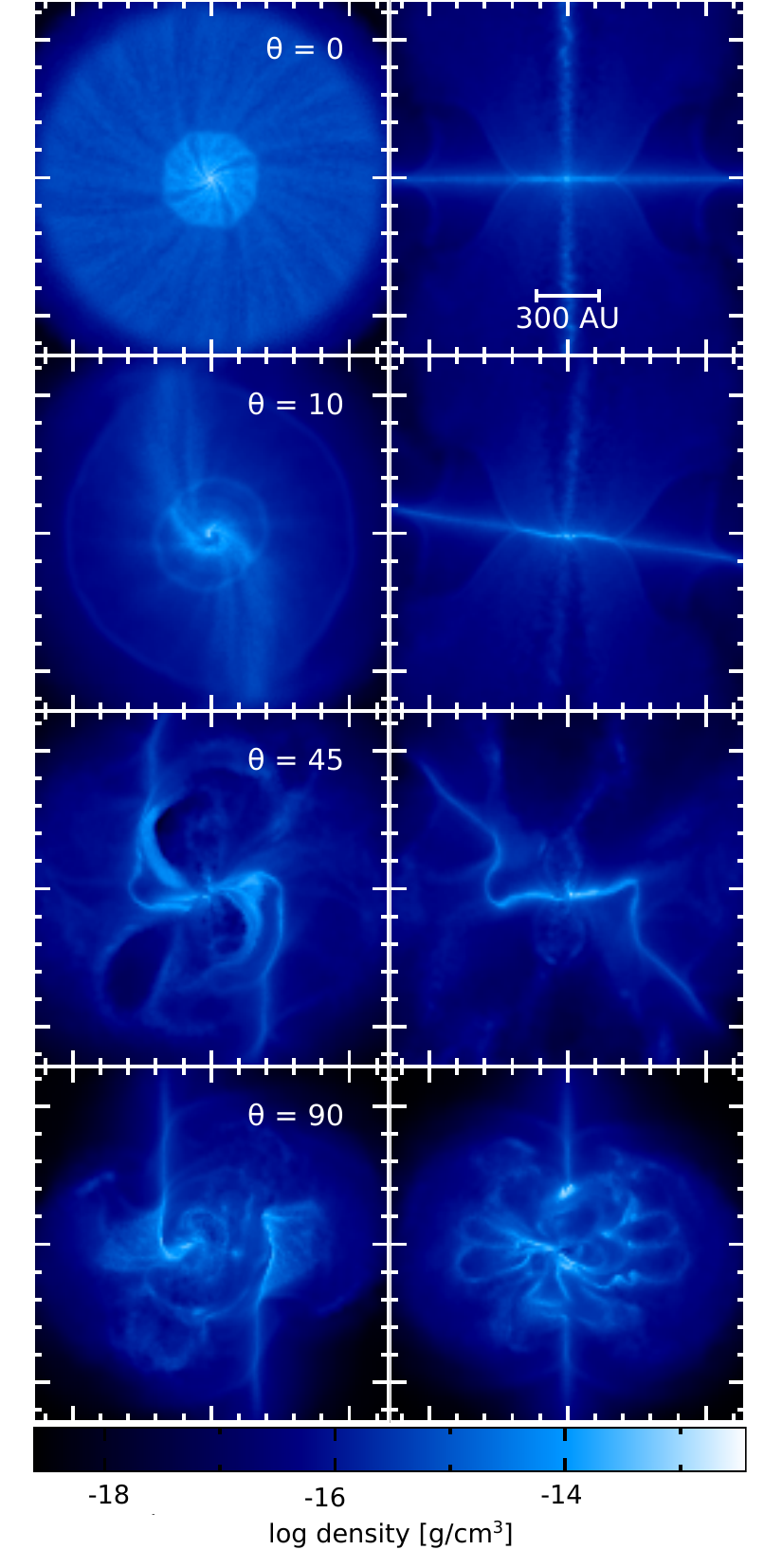}

\caption{Cross-sections of density in the $x$-$y$ plane (left column) and $z$-$y$ plane (right column) at $t$ = 27870 yrs for the same values of $\vartheta$ as in \cref{fig:modJ}. The uniform pseudo-disc, perpendicular to the rotation axis, can be clearly seen in the aligned case, as can the formation of a bar like structure for $\vartheta = 10\degree$. The \diff{jets in the} lower two pairs of plots are, as expected given the complicated velocity profile, completely disrupted.\label{fig:rhocut}}
\end{figure}

We observe in \cref{fig:modJ} that as the magnetic field geometry near the sink becomes very complicated bubbles of material driven by magnetic pressure form and disrupt the accretion of matter into a disc. For $\vartheta \ge{} 60\degree$ this is sufficient to suppress the \diff{formation of a disc and outflow altogether}, whilst for shallower angles the outflow simply becomes more diffuse and less structured. Compared to the aligned case, even seting $\vartheta \leq{} 10\degree$ has an effect on the pseudo-disc (\cref{fig:rhocut}). In and of itself, such a structure should not prevent a collimated jet being produced -- and indeed one is seen in both the $10$ and $20\degree$ models with a similar velocity to the simpler aligned case. \Cref{fig:10degzoom} shows the evolution of the jet for $\vartheta{} = 10\degree{}$. Initially, the most central region of the pseudo-disc aligns perpendicularly with the rotation axis, and consequently a jet is produced parallel to this axis, whilst the outer regions remain aligned perpendicular to the field axis. As the system evolves, the central portion warps, this causes the outflow to realign with the field axis, causing the kink seen in \cref{fig:fig1} at the extremities of the jet.

\begin{figure}
\centering{}
\includegraphics{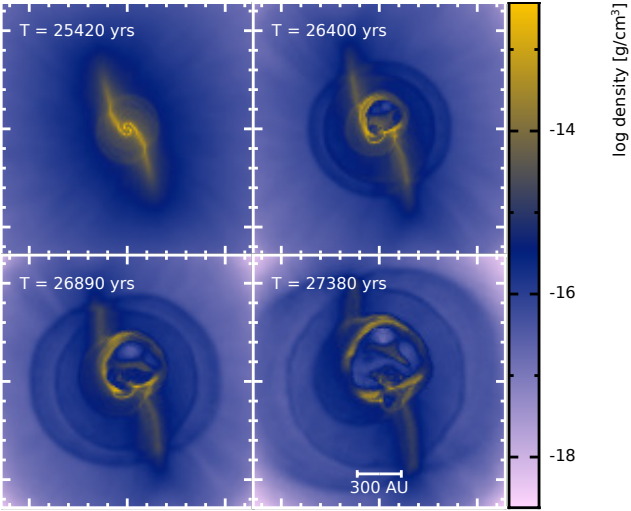}

\caption{Density cross-sections for $\vartheta = 20\degree$. The effect of the complicated field geometry in expelling material from around the protostar, and thus distrupting the collimated jet, can be clearly seen.\label{fig:rhocut_20degxy}}
\end{figure}

In \cref{fig:rhocut_20degxy} we plot a series of density cross-sections for the $\vartheta = 20\degree$ model. The same disc warping seen at $10\degree$ is present, however, unlike the shallower angle the jet produced here is disrupted within 700 years of forming. The outflow continues, as seen in \cref{fig:fig1}, and has a profile broadly similar to shallower angles but without a central jet. This loss of collimation appears to be the result of the formation of a bubble of material near the protostar 
which pushes material away from the core and thereby prevent the formation of a high velocity \diff{region of the} pseudo-disc which can collimate an outflow. This is not a true accretion disc with a Keplerian velocity profile and is simply a result of rotational and magnetic forces forcing material into a disc-like structure. Whilst magnetic bubbles of this nature have been seen before (see, e.g. \citet{2011ApJ...742...10Z}, \citet{2012ApJ...757...77K}) it is unclear whether this is a real effect or due to the lack of any physical resistivity or other non-ideal MHD effects (e.g. ambipolar diffusion or the Hall effect) in our model. Resistivity may allow for magnetic reconnection and hence the conversion of magnetic energy into thermal energy and may act to stabilise the accretion structures in this case. 

\begin{figure}
\centering{}
\includegraphics{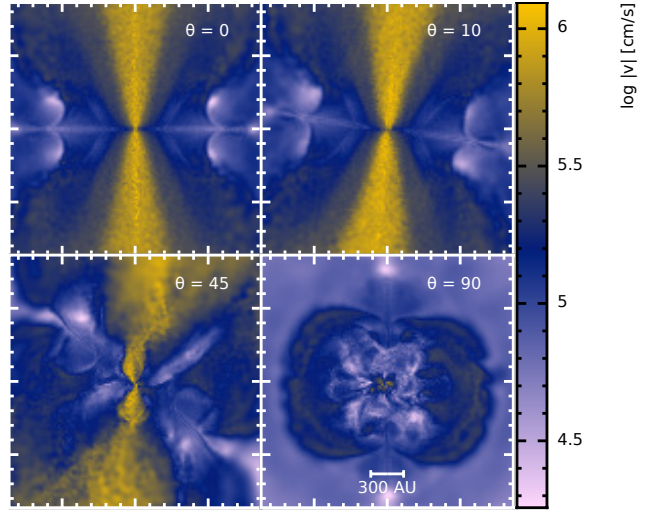}

\caption{Cross-sections of $|\mathbf{v}|$ at T = 27870 yrs (i.e. late enough that the magnetic effects seen in \cref{fig:modJ} will have been able to disrupt the outflow). The two upper plots show the strongly collimated outflows seen for low values
of $\vartheta$ whilst the $\vartheta = 45$ and $90\degree$ plots show the more complicated, diffuse outflows which at steeper angles essentially cease altogether.\label{fig:modV}}
\end{figure}

A similar effect can be seen for $\vartheta = 45\degree$, however in this case the effect is so pronounced that the outflow is completely disrupted and very large, broadly symmetrical `puffy' outflow can be seen. Comparing the shallower angles with this model in \cref{fig:modV} shows that the actual velocity profile consists of two zones: a narrow region just above and below the sink that is aligned with the rotation axis, and a diffuse yet still fast zone further out which broadly follows the field axis. 
The alignment of the outflow initially with the rotation axis is caused by the central region of the pseudo-disc being aligned (as shown in \cref{fig:rhocut}) perpendicular to the rotation axis. As the outflow moves away, it will be acted on by magnetic force which will re-orient it to align with the field axis so that the fluid moves along the field lines.
Unlike for $\vartheta = 20\degree$ and $60\degree$, the bubble structures seen at this angle remain generally symmetrical; a likely explanation for this is simply that for $45\degree$ the magnitude of the field is identical in both the $x$ and $z$ direction and consequently there is nothing to create an asymmetry.

\diff{We note that the jets seen in the more aligned cases (ca.} $8~\text{km~s}^{-1}$\diff{ ) are significantly slower than those observed in observations of Herbig-Haro objects (where velocities of }$>100~\text{km~s}^{-1}$\diff{ are commonly seen) and are also slower than the fastest velocities observed in }\citet{2014MNRAS.437...77B}\diff{. This is caused by the relatively large sink radius which is orders of magnitude larger than the stellar core, and the consequent relatively slow maximum Keplerian velocity (in this case at 1 AU). Since jet velocity is proportional to the velocity of the antecedent accretion disc }\citep{2003MNRAS.339.1223P}\diff{ this naturally produces a slower jet.}

\begin{figure}
\centering{}
\includegraphics{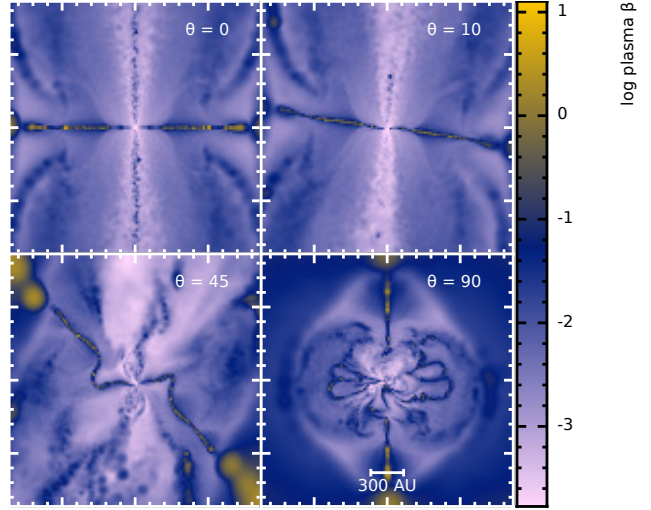}

\caption{Plasma $\beta$ cross-sections for $\vartheta = 0\degree$ and $90\degree$. In the aligned case, a pressure supported pseudo-disc (aligned perpendicularly to the field) is present, whilst for the misaligned case it is completely disrupted by magnetic pressure several hundred AU away.\label{fig:plasbeta}}
\end{figure}

The lack of any outflow, as opposed to a simple puffing out of the core region, in the $\vartheta = 90\degree$ model is expected since a field geometry this extreme will naturally lead to a very complicated distribution of field and mass near the sink. In \cref{fig:plasbeta} we compare the plasma $\beta$ in the aligned and most non-aligned cases. This clearly shows that bubbles of material, driven by magnetic pressure disrupt the accretion disc whilst in the aligned case this hydrodynamically dominated disc persists all the way to the centre and ultimately the accretion region. As seen previously, \cref{fig:modV} shows that the velocities in this case are substantially lower than for the aligned model (the maximum outflow velocity is approximately $2~\rmn{km~s^{-1}}$, compared to $8~\rmn{km~s^{-1}}$ when $\vartheta = 0 \degree$) providing further evidence that magnetic pressure is influencing the collapse. 

We obtain generally similar morphologies to those seen in \citet{2010MNRAS.409L..39C}, where the outflows become increasingly disrupted as $\vartheta$ is increased. For example, at $\vartheta = 45\degree$ a puffy outflow with no distinct jet is obtained in both models, and at $\approx{} 27000$ yrs this outflow is $\approx{} 2000-3000$ AU in size; and at higher angles the outflow is suppressed until being essentially extinguished at $\vartheta = 90\degree$. Both models also agree for $\vartheta = 20\degree$ until $\approx 22000$ yrs when \citet{2010MNRAS.409L..39C} stop, however, we observe that the jet is subsequently disrupted and replaced with a diffuse outflow.

\subsection{Accretion}

\begin{figure}
\centering{}
\includegraphics{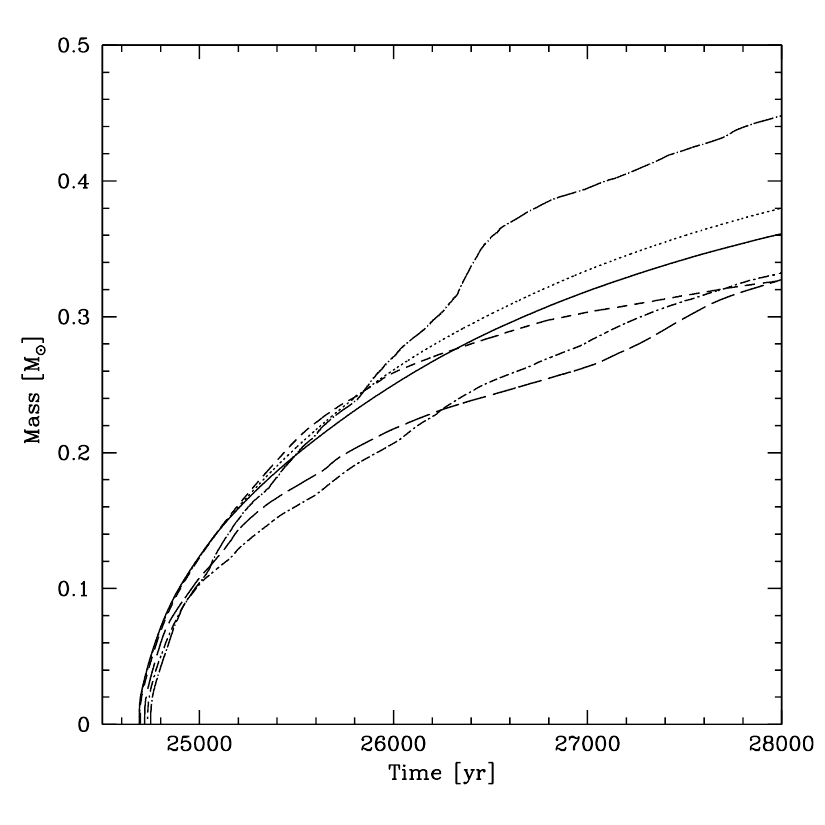}

\caption{Accretion of mass by the sink particle for $\vartheta = 0\degree$ (solid), $10\degree$ (dotted), $20\degree$ (short dashes), $45\degree$ (long dashes), $60\degree$ (dot -- short dashes), and $90\degree$ (dot -- long dashes). The $10\degree$ model accretes slightly faster than the aligned model; all steeper angles except $90\degree$ accrete less material, although $20\degree$ initially follows $10\degree$. There is a sharp `knee' in the $90\degree$ model at approximately 28000 yrs.\label{fig:fig3}}
\end{figure}

In the early part of the simulation, after the sink is inserted, we observe very rapid accretion. This rate decreases over time, both due to matter being expelled from the core by outflows and also due to the dynamics of the collapse \citep{1993ApJ...416..303F},
however, the eventual accretion rate does depend on the value of $\vartheta$ used.

\Cref{fig:fig3} shows the mass accreted by the sink particle for each model as a function of time. We would assume that for ever increasing values of $\vartheta$, the accretion rate will fall as the complicated field effects (ranging from loops of material to the more extreme $90\degree$ model) push material away from the sink particle. Between $\vartheta \approx{} 20\degree$ and $\vartheta \approx{} 60\degree$
the accretion rate  does indeed fall for steeper angles. However this trend markedly reverses for the steepest angles, in particular the rate for $90\degree$ is substantially faster and is observed to increase sharply at a `knee' rather than decrease. This is a counter-intuitive result given the structure of the cloud core and the very low plasma $\beta$ in this regime. This differs substantially from the result \citet{2010MNRAS.409L..39C} who found that there was no regime in which the accretion rate fell. \citet{2010MNRAS.409L..39C} developed an analytical model (from the model for an isothermal collapsing sphere in \citet{1977ApJ...218..834H}) whereby
\begin{equation}
M_{\rmn{core}}\left(t\right) = \tau{}_{\rmn{ae}} ~\dot{M}_{\rmn{inf}}\left(1 - \rmn{exp~} \frac{-t~}{\tau{}_{\rmn{ae}}} \right) \text{~,} \label{eqn:CH1}
\end{equation}
where $\dot{M}_{\rmn{inf}}$ is a constant determined from the sound speed of the medium and
\begin{equation}
\tau{}_{\rmn{ae}} \propto{} \frac{1}{\rmn{cos}~\vartheta{}} \text{~.} \label{eqn:CH2}
\end{equation}
\Cref{eqn:CH1,eqn:CH2} will produce ever faster accretion rates as $\vartheta$ increases without the decrease we see for values of $20\degree \leq{} \vartheta < 90\degree$ - and clearly breaks down when $\vartheta{} = 90\degree$. This is simply a result of the assumptions made, i.e. that the sphere collapses held up only by magnetic pressure (and by material being removed by the outflow). We do obtain an increased accretion rate in the $10\degree$ model (and initially, before magnetic bubbles distrupt the pseudo-disc, in the $20\degree$ model) which lends some support to this hypothesis. In contrast, the complicated bubble structures seen at larger angles disrupt the accretion process and therefore cause a reduction in the core mass. At the largest values of $\vartheta$, the outflow is so suppressed, however, that the accretion rate actually increases. 

\begin{figure}
\centering{}
\includegraphics{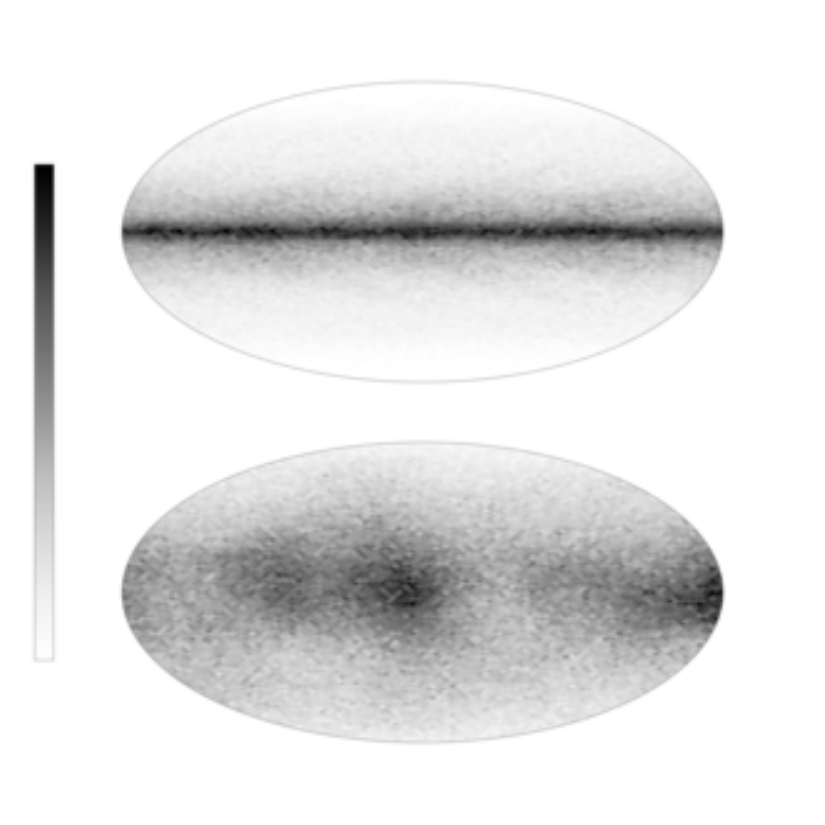}

\caption{Latitude-longitude maps of particle accretion for the first 4,000 after a sink is inserted. The upper plot is for the $\vartheta = 0\degree$ case and the lower plot for $\vartheta = 90\degree$. \diff{Darker} colours, which denote a greater number of particles accreted show that for the aligned case the accretion is clearly disc-linked, and conversely that the misaligned accretion covers a substantially larger solid angle. \diff{The color scale is normalised so that white represents a pixel with no accreted particles and black represents the pixel with the highest number of accreted particles.} \label{fig:fig4}}
\end{figure}

We see in \cref{fig:fig4} one of the causes of this increase in accretion rate at $\vartheta = 90\degree$. For strongly aligned fields and rotation axes, the accretion process can only happen along the edge of the disc. This remains true even as the disc itself is disrupted by magnetic effects -- in essence, rather than being a constant equatorial line, material is accreted at the edges of the bubbles and other disturbances in the pseudo-disc. For much larger angles, the solid angle over which accretion can occur is much larger - whilst the rotational forces are trying to hold the material into a disc-like structure, the magnetic bubbles formed completely disrupt this. There is a still a general preference to accrete material along the equator of the sink, rather than the poles because the material is still spinning. 

\section{Conclusion}
\label{sec:conc}

We have devised a small modification to the equations of SPMHD, implementing the induction equation and anisotropic force equation using an average smoothing length formalism, to eliminate an instability that has hitherto prevented work using smaller sink particles and misaligned magnetic fields in SPMHD models of collapsing magnetised cloud cores. We produced six models of these cloud cores, with angles between the field and rotation axes ranging from $0\degree$ (aligned) to $90\degree$. In each simulation, we were able to follow the collapse from an initially cold sphere for over 26,000 years 
to the formation of a dense core, accretion disc and outflows. 

We observe that the nature of the outflows observed varies strongly with the initial field geometry. For the aligned, i.e. $\vartheta = 0\degree$, model we observe an outflow similar to that obtained in previous work \citep{2012MNRAS.423L..45P} albeit with a slightly faster collimated jet due to the smaller sink radius, with the characteristic jet and envelope seen in previous work. For misaligned fields, we are able to divide the nature of the outflow into three regimes: for very shallow angles ($\ge{} 10\degree$) a collimated jet is produced; for moderate angles ($20\degree - 45\degree$) an initial collimated jet is rapidly disrupted by the increasingly wound up magnetic field which produces large magnetic bubbles and the outflow becomes much more diffuse and unstructured. Finally, for the steepest angles the outflow is substantially suppressed and becomes more spherical.

\section*{Acknowledgments}

BTL and MRB thank Monash University for their hospitality during April-July 2014, when the majority of this work was completed and Joe Monaghan for his sage advice.

This work was supported by the European Research Council under the European Community's Seventh Framework Programme (FP7/2007-2013 Grant Agreement No.
339248). BTL also acknowledges support from an STFC Studentship and Long Term Attachment grant.
MRB's visit to Monash was funded by an International Collaboration Award from the Australian Research Council (ARC) under the Discovery Project scheme grant DP130102078. DJP acknowledges funding from the ARC via DP130102078 and FT130100034.

The calculations for this paper were performed on the DiRAC Complexity machine, jointly funded by STFC and the Large Facilities Capital Fund of BIS,
and the University of Exeter Supercomputer, a DiRAC Facility jointly funded by STFC, the Large Facilities Capital Fund of BIS and the University of
Exeter.

Rendered plots were produced using the \texttt{SPLASH} \citep{2007PASA...24..159P} visualisation programme. BTL would like to especially thank Elisabeth Matthews for assistance in producing aesthetically pleasing plots.

\renewcommand{\refname}{\textsc{References}}
\bibliography{Outflows_misaligned_fields}

\begin{thebibliography}{}

\bibitem[\protect\citeauthoryear{{Bate}}{{Bate}}{2009}]{2009MNRAS.392..590B}
{Bate} M.~R.,  2009, \mnras, 392, 590

\bibitem[\protect\citeauthoryear{{Bate}}{{Bate}}{2012}]{2012MNRAS.419.3115B}
{Bate} M.~R.,  2012, \mnras, 419, 3115

\bibitem[\protect\citeauthoryear{{Bate}, {Bonnell} \& {Price}}{{Bate}
  et~al.}{1995}]{1995MNRAS.277..362B}
{Bate} M.~R.,  {Bonnell} I.~A.,    {Price} N.~M.,  1995, \mnras, 277, 362

\bibitem[\protect\citeauthoryear{{Bate} \& {Burkert}}{{Bate} \&
  {Burkert}}{1997}]{1997MNRAS.288.1060B}
{Bate} M.~R.,  {Burkert} A.,  1997, \mnras, 288, 1060

\bibitem[\protect\citeauthoryear{{Bate}, {Tricco} \& {Price}}{{Bate}
  et~al.}{2014}]{2014MNRAS.437...77B}
{Bate} M.~R.,  {Tricco} T.~S.,    {Price} D.~J.,  2014, \mnras, 437, 77

\bibitem[\protect\citeauthoryear{{Benz}, {Cameron}, {Press} \& {Bowers}}{{Benz}
  et~al.}{1990}]{1990ApJ...348..647B}
{Benz} W.,  {Cameron} A.~G.~W.,  {Press} W.~H.,    {Bowers} R.~L.,  1990, \apj,
  348, 647

\bibitem[\protect\citeauthoryear{{B{\o}rve}, {Omang} \& {Trulsen}}{{B{\o}rve}
  et~al.}{2001}]{2001ApJ...561...82B}
{B{\o}rve} S.,  {Omang} M.,    {Trulsen} J.,  2001, \apj, 561, 82

\bibitem[\protect\citeauthoryear{{Ciardi} \& {Hennebelle}}{{Ciardi} \&
  {Hennebelle}}{2010}]{2010MNRAS.409L..39C}
{Ciardi} A.,  {Hennebelle} P.,  2010, \mnras, 409, L39

\bibitem[\protect\citeauthoryear{{Crutcher}}{{Crutcher}}{2012}]{2012ARA&A..50...29C}
{Crutcher} R.~M.,  2012, \araa, 50, 29

\bibitem[\protect\citeauthoryear{{Crutcher}, {Troland}, {Goodman}, {Heiles},
  {Kazes} \& {Myers}}{{Crutcher} et~al.}{1993}]{1993ApJ...407..175C}
{Crutcher} R.~M.,  {Troland} T.~H.,  {Goodman} A.~A.,  {Heiles} C.,  {Kazes}
  I.,    {Myers} P.~C.,  1993, \apj, 407, 175

\bibitem[\protect\citeauthoryear{{Dedner}, {Kemm}, {Kr{\"o}ner}, {Munz},
  {Schnitzer} \& {Wesenberg}}{{Dedner} et~al.}{2002}]{2002JCoPh.175..645D}
{Dedner} A.,  {Kemm} F.,  {Kr{\"o}ner} D.,  {Munz} C.-D.,  {Schnitzer} T.,
  {Wesenberg} M.,  2002, Journal of Computational Physics, 175, 645

\bibitem[\protect\citeauthoryear{{Donati}, {Skelly}, {Bouvier}, {Gregory},
  {Grankin}, {Jardine}, {Hussain}, {M{\'e}nard}, {Dougados}, {Unruh},
  {Mohanty}, {Auri{\`e}re}, {Morin}, {Far{\`e}s} \& {MAPP
  Collaboration}}{{Donati} et~al.}{2010}]{2010MNRAS.409.1347D}
{Donati} J.-F.,  {Skelly} M.~B.,  {Bouvier} J.,  {Gregory} S.~G.,  {Grankin}
  K.~N.,  {Jardine} M.~M.,  {Hussain} G.~A.~J.,  {M{\'e}nard} F.,  {Dougados}
  C.,  {Unruh} Y.,  {Mohanty} S.,  {Auri{\`e}re} M.,  {Morin} J.,  {Far{\`e}s}
  R.,    {MAPP Collaboration} 2010, \mnras, 409, 1347

\bibitem[\protect\citeauthoryear{{Fehlberg}}{{Fehlberg}}{1969}]{RK4NASATR}
{Fehlberg} E.,  1969, {NASA Technical Report}, R-315

\bibitem[\protect\citeauthoryear{{Foster} \& {Chevalier}}{{Foster} \&
  {Chevalier}}{1993}]{1993ApJ...416..303F}
{Foster} P.~N.,  {Chevalier} R.~A.,  1993, \apj, 416, 303

\bibitem[\protect\citeauthoryear{{Gingold} \& {Monaghan}}{{Gingold} \&
  {Monaghan}}{1977}]{1977MNRAS.181..375G}
{Gingold} R.~A.,  {Monaghan} J.~J.,  1977, \mnras, 181, 375

\bibitem[\protect\citeauthoryear{{Goodman}, {Benson}, {Fuller} \&
  {Myers}}{{Goodman} et~al.}{1993}]{1993ApJ...406..528G}
{Goodman} A.~A.,  {Benson} P.~J.,  {Fuller} G.~A.,    {Myers} P.~C.,  1993,
  \apj, 406, 528

\bibitem[\protect\citeauthoryear{{Hull}, {Plambeck}, {Bolatto}, {Bower},
  {Carpenter}, {Crutcher}, {Fiege}, {Franzmann}, {Hakobian} \& {Heiles}}{{Hull}
  et~al.}{2013}]{2013ApJ...768..159H}
{Hull} C.~L.~H.,  {Plambeck} R.~L.,  {Bolatto} A.~D.,  {Bower} G.~C.,
  {Carpenter} J.~M.,  {Crutcher} R.~M.,  {Fiege} J.~D.,  {Franzmann} E.,
  {Hakobian} N.~S.,    {Heiles} C.,  2013, \apj, 768, 159

\bibitem[\protect\citeauthoryear{{Hunter}}{{Hunter}}{1977}]{1977ApJ...218..834H}
{Hunter} C.,  1977, \apj, 218, 834

\bibitem[\protect\citeauthoryear{{Krasnopolsky}, {Li}, {Shang} \&
  {Zhao}}{{Krasnopolsky} et~al.}{2012}]{2012ApJ...757...77K}
{Krasnopolsky} R.,  {Li} Z.-Y.,  {Shang} H.,    {Zhao} B.,  2012, \apj, 757, 77

\bibitem[\protect\citeauthoryear{{Larson}}{{Larson}}{1969}]{1969MNRAS.145..271L}
{Larson} R.~B.,  1969, \mnras, 145, 271

\bibitem[\protect\citeauthoryear{{Lucy}}{{Lucy}}{1977}]{1977AJ.....82.1013L}
{Lucy} L.~B.,  1977, \aj, 82, 1013

\bibitem[\protect\citeauthoryear{{Mac Low} \& {Klessen}}{{Mac Low} \&
  {Klessen}}{2004}]{2004RvMP...76..125M}
{Mac Low} M.-M.,  {Klessen} R.~S.,  2004, Reviews of Modern Physics, 76, 125

\bibitem[\protect\citeauthoryear{{Machida}, {Inutsuka} \&
  {Matsumoto}}{{Machida} et~al.}{2008}]{2008ApJ...676.1088M}
{Machida} M.~N.,  {Inutsuka} S.-i.,    {Matsumoto} T.,  2008, \apj, 676, 1088

\bibitem[\protect\citeauthoryear{{Monaghan}}{{Monaghan}}{1985}]{1985JCoPh..60..253M}
{Monaghan} J.~J.,  1985, Journal of Computational Physics, 60, 253

\bibitem[\protect\citeauthoryear{{Monaghan}}{{Monaghan}}{1997}]{1997JCoPh.136..298M}
{Monaghan} J.~J.,  1997, Journal of Computational Physics, 136, 298

\bibitem[\protect\citeauthoryear{{Morris} \& {Monaghan}}{{Morris} \&
  {Monaghan}}{1997}]{1997JCoPh.136...41M}
{Morris} J.~P.,  {Monaghan} J.~J.,  1997, Journal of Computational Physics,
  136, 41

\bibitem[\protect\citeauthoryear{{Mouschovias} \& {Spitzer} Jr.}{{Mouschovias}
  \& {Spitzer}}{1976}]{1976ApJ...210..326M}
{Mouschovias} T.~C.,  {Spitzer} Jr. L.,  1976, \apj, 210, 326

\bibitem[\protect\citeauthoryear{{Nakano} \& {Nakamura}}{{Nakano} \&
  {Nakamura}}{1978}]{1978PASJ...30..671N}
{Nakano} T.,  {Nakamura} T.,  1978, \pasj, 30, 671

\bibitem[\protect\citeauthoryear{{Phillips} \& {Monaghan}}{{Phillips} \&
  {Monaghan}}{1985}]{1985MNRAS.216..883P}
{Phillips} G.~J.,  {Monaghan} J.~J.,  1985, \mnras, 216, 883

\bibitem[\protect\citeauthoryear{{Price}}{{Price}}{2007}]{2007PASA...24..159P}
{Price} D.~J.,  2007, \pasa, 24, 159

\bibitem[\protect\citeauthoryear{{Price}}{{Price}}{2012}]{2012JCoPh.231..759P}
{Price} D.~J.,  2012, Journal of Computational Physics, 231, 759

\bibitem[\protect\citeauthoryear{{Price} \& {Bate}}{{Price} \&
  {Bate}}{2008}]{2008MNRAS.385.1820P}
{Price} D.~J.,  {Bate} M.~R.,  2008, \mnras, 385, 1820

\bibitem[\protect\citeauthoryear{{Price} \& {Bate}}{{Price} \&
  {Bate}}{2009}]{2009MNRAS.398...33P}
{Price} D.~J.,  {Bate} M.~R.,  2009, \mnras, 398, 33

\bibitem[\protect\citeauthoryear{{Price} \& {Monaghan}}{{Price} \&
  {Monaghan}}{2004}]{2004MNRAS.348..139P}
{Price} D.~J.,  {Monaghan} J.~J.,  2004, \mnras, 348, 139

\bibitem[\protect\citeauthoryear{{Price} \& {Monaghan}}{{Price} \&
  {Monaghan}}{2005}]{2005MNRAS.364..384P}
{Price} D.~J.,  {Monaghan} J.~J.,  2005, \mnras, 364, 384

\bibitem[\protect\citeauthoryear{{Price} \& {Monaghan}}{{Price} \&
  {Monaghan}}{2007}]{2007MNRAS.374.1347P}
{Price} D.~J.,  {Monaghan} J.~J.,  2007, \mnras, 374, 1347

\bibitem[\protect\citeauthoryear{{Price}, {Pringle} \& {King}}{{Price}
  et~al.}{2003}]{2003MNRAS.339.1223P}
{Price} D.~J.,  {Pringle} J.~E.,    {King} A.~R.,  2003, \mnras, 339, 1223

\bibitem[\protect\citeauthoryear{{Price}, {Tricco} \& {Bate}}{{Price}
  et~al.}{2012}]{2012MNRAS.423L..45P}
{Price} D.~J.,  {Tricco} T.~S.,    {Bate} M.~R.,  2012, \mnras, 423, L45

\bibitem[\protect\citeauthoryear{{Stephens}, {Looney}, {Kwon},
  {Fern{\'a}ndez-L{\'o}pez}, {Hughes}, {Mundy}, {Crutcher}, {Li} \&
  {Rao}}{{Stephens} et~al.}{2014}]{2014Natur.514..597S}
{Stephens} I.~W.,  {Looney} L.~W.,  {Kwon} W.,  {Fern{\'a}ndez-L{\'o}pez} M.,
  {Hughes} A.~M.,  {Mundy} L.~G.,  {Crutcher} R.~M.,  {Li} Z.-Y.,    {Rao} R.,
  2014, \nat, 514, 597

\bibitem[\protect\citeauthoryear{{Swegle}, {Hicks} \& {Attaway}}{{Swegle}
  et~al.}{1995}]{1995JCoPh.116..123S}
{Swegle} J.~W.,  {Hicks} D.~L.,    {Attaway} S.~W.,  1995, Journal of
  Computational Physics, 116, 123

\bibitem[\protect\citeauthoryear{{Tricco} \& {Price}}{{Tricco} \&
  {Price}}{2012}]{2012JCoPh.231.7214T}
{Tricco} T.~S.,  {Price} D.~J.,  2012, Journal of Computational Physics, 231,
  7214

\bibitem[\protect\citeauthoryear{{Tricco} \& {Price}}{{Tricco} \&
  {Price}}{2013}]{2013MNRAS.436.2810T}
{Tricco} T.~S.,  {Price} D.~J.,  2013, \mnras, 436, 2810

\bibitem[\protect\citeauthoryear{{Zhao}, {Li}, {Nakamura}, {Krasnopolsky} \&
  {Shang}}{{Zhao} et~al.}{2011}]{2011ApJ...742...10Z}
{Zhao} B.,  {Li} Z.-Y.,  {Nakamura} F.,  {Krasnopolsky} R.,    {Shang} H.,
  2011, \apj, 742, 10

\end{thebibliography}

\bsp

\label{lastpage}

\end{document}